\documentclass[conference]{IEEEtran}
\IEEEoverridecommandlockouts
\usepackage{cite}
\usepackage{amsmath,amssymb,amsfonts}
\usepackage{algorithmic}
\usepackage{graphicx}
\usepackage{textcomp}
\usepackage{hyperref}
\usepackage{fancyhdr}
\usepackage{xcolor}
\def\BibTeX{{\rm B\kern-.05em{\sc i\kern-.025em b}\kern-.08em
    T\kern-.1667em\lower.7ex\hbox{E}\kern-.125emX}}

\fancypagestyle{FirstPage}{
\lfoot{COPYRIGHT \copyright 2023 IEEE.  Personal use of this material is permitted.  Permission from IEEE must be obtained for all other uses, in any current or future media, including reprinting/republishing this material for advertising or promotional purposes, creating new collective works, for resale or redistribution to servers or lists, or reuse of any copyrighted component of this work in other works. Accepted at 2023 IEEE International Conference on Medical Artificial Intelligence (MedAI)} 
}

\begin{document}

\title{Path Signature Representation of Patient-Clinician Interactions as a Predictor for Neuropsychological Tests Outcomes in Children: A Proof of Concept
}
\author{\IEEEauthorblockN{\;\;\;\;\;\;\;1\textsuperscript{st}Giulio Falcioni\;\;\;\;\;\;\;\;\;\;\;\;\;\;}
\IEEEauthorblockA{\textit{\;\;\;\;\;\;\;The University of Edinburgh\;\;\;\;\;\;\;\;\;\;\;\;}\\
\;\;\;\;\;\;\;Edinburgh, United Kingdom \;\;\;\;\;\;\;\;\;\;\;\;\\\
\;\;\;\;\;\;giulio.falcioni@ed.ac.uk\;\;\;\;\;\;\;\;\;\;\;\\}
\and
\IEEEauthorblockN{\;\;\;\;\;\;\;2\textsuperscript{nd} Alexandra Georgescu\;\;\;\;\;\;\;\;\;\;}
\IEEEauthorblockA{\textit{Thymia Limited\;\;\;\;\;\;\;\;\;\;}\\
London, United Kingdom\;\;\;\;\;\;\;\;\;\;\\
alexandra@thymia.ai\;\;\;\;\;\;\;\;\;\;}
\and
\IEEEauthorblockN{3\textsuperscript{rd} Emilia Molimpakis\;\;\;\;\;\;\;\;\;\;\;}
\IEEEauthorblockA{\textit{Thymia Limited\;\;\;\;\;\;\;\;\;\;\;\;\;}\\
London, United Kingdom\;\;\;\;\;\;\;\;\;\;\;\\
emilia@thymia.ai\;\;\;\;\;\;\;\;\;\;\;\;\;\;}
\and
\IEEEauthorblockN{\;\;\;\;\;\;\;\;\;\;4\textsuperscript{th} Lev Gottlieb\;\;\;\;\;\;\;\;\;\;}
\IEEEauthorblockA{\textit{\;\;\;\;\;\;\;\;\;\;The Integrated Clinic\;\;\;\;\;\;\;\;\;\;} \\
\;\;\;\;\;\;\;\;\;\;Los Angeles, United States\;\;\;\;\;\;\;\;\;\;\\
\;\;\;\;\;\;\;\;\;\;lev@theintegratedclinic.com\;\;\;\;\;\;\;\;\;\;}
\and
\IEEEauthorblockN{5\textsuperscript{th} Taylor Kuhn}
\IEEEauthorblockA{\textit{The Integrated Clinic} \\
Los Angeles, United States\\
taylor@theintegratedclinic.com}
\and
\IEEEauthorblockN{6\textsuperscript{th} Stefano Goria}
\IEEEauthorblockA{\textit{Thymia Limited} \\
London, United Kingdom \\
stefano@thymia.ai}
}


\maketitle

\begin{abstract}
This research report presents a proof-of-concept study on the application of machine learning techniques to video and speech data collected during diagnostic cognitive assessments of children with a neurodevelopmental disorder. The study utilised a dataset of 39 video recordings, capturing extensive sessions where clinicians administered, among other things, four cognitive assessment tests. From the first 40 minutes of each clinical session, covering the administration of the Wechsler Intelligence Scale for Children (WISC-5), we extracted head positions and speech turns of both clinician and child. Despite the limited sample size and heterogeneous recording styles, the analysis successfully extracted path signatures as features from the recorded data, focusing on patient-clinician interactions. Importantly, these features quantify the interpersonal dynamics of the assessment process (dialogue and movement patterns). Results suggest that these features exhibit promising potential for predicting all cognitive tests scores of the entire session length and for prototyping a predictive model as a clinical decision support tool. Overall, this proof of concept demonstrates the feasibility of leveraging machine learning techniques for clinical video and speech data analysis in order to potentially enhance the efficiency of cognitive assessments for neurodevelopmental disorders in children.
\end{abstract}


\section{Introduction}
\thispagestyle{FirstPage}
Neuropsychological assessments are part of the gold standard in diagnostic assessment for neurodevelopmental disorders, such as Attention Deficit Hyperactivity Disorder (ADHD), Autism Spectrum Disorder (ASD) and Learning Disorders. They involve an extensive battery of cognitive tests designed to assess various domains, such as attention, memory, executive function, social and language abilities. They provide valuable insights into an individual's cognitive functioning, aiding clinicians in making accurate diagnoses and formulating appropriate treatment plans. 

With increasing prevalence rates of neurodevelopmental disorders, the number of referrals for diagnosis has also skyrocketed \cite{maenner2020prevalence},\cite{mccarthy2012epidemiology}. This increase can be attributed to factors such as broader diagnostic criteria and heightened awareness among professionals and the public. Assessments are, however, very time- and labour-intensive, leading to lengthy waiting periods for individuals seeking diagnostic clarification. With the increasing demand on their services, diagnosis providers have been overwhelmed, leading in some cases to the closure of already long waiting lists (e.g.\cite{Topping.2023}). Unfortunately, this leads to delays in individuals accessing essential support, interventions and academic adjustments, with detrimental effects on the overall academic environment and significant impacts on the individual's development. Delays in accessing a diagnosis also contribute to various mental health comorbidities. In ADHD, for example, comorbidities include depression (30\%), anxiety (37.9\%), and behavioral disorders (31\%, specifically oppositional defiant disorder) \cite{mohammadi2021prevalence},\cite{angold1999comorbidity}.
 
With an overburdened healthcare system involving a narrow bottleneck funnel to assessment leading to prolonged delays between referrals, assessment, diagnosis, treatment planning and receipt of both treatment and academic accommodations, children and adolescents with neurodevelopmental disorders like ADHD and ASD are not receiving the help they need when they need it. There is an urgent need for more efficient assessment processes to shorten and facilitate both access to the pathway of diagnostic assessment and its completion. 

This need could be filled by the application of machine learning (ML) techniques, as they offer an automated and accurate way of detecting symptoms and characteristics of various disorders or conditions. The application of machine learning techniques in screening ADHD, ASD and other neurodevelopmental and learning conditions has gained considerable attention in recent years (e.g. \cite{nash2022machine}). For example, some attempts have been made to expand on traditional gold-standard assessment tools with parent-report questionnaires and remote clinical ratings of home videos uploaded by parents. Such tools have shown early evidence of providing accurate diagnosis of autism \cite{kanne2018screening}. Furthermore, ML approaches applied to informant and self-report questionnaires have been shown to accurately differentiate between autistic children and those with ADHD\cite{duda2017crowdsourced}. A drawback of these methods, however, is that they rely on subjective ratings and/or semi-structured assessments that are still time- and labour-intensive whilst also depending on the clinician's experience and the language they are administered in.

A more objective ML approach has been to use automated speech, eye and/or body movement analyses. Unlike subjective parent-reported questionnaires, self-reports or assessments reliant solely on clinical experience, signals from movement or speech offer objective markers as a measure for analysis. For example, this can include accelerometer or pupil data for ADHD \cite{park2023machine},\cite{das2021robust} or voice and interpersonal coordination data for ASD \cite{rybner2022vocal},\cite{georgescu2019machine}. While presenting promising results, this approach is heavily dependant on identifying gaze, movement and speech markers that are not only shared among the diverse phenotypes of these disorders but, crucially, are simultaneously distinctive of each specific condition. This is essential for developing clinical decision support tools that ensure not only high sensitivity but also specificity, contributing to more reliable referrals at the screening stage and fewer misdiagnoses at the diagnostic stage. However, most ML approaches do not consider the inclusion of control groups other than typically developed individuals and children, hence making it difficult to adequately identify behavioural markers and/or validate algorithms for clinical use.

We present an alternative approach that seeks to make neuropsychological assessment in neurodevelopmental disorders more efficient and ultimately more equitable. This research report focuses on the utilization of machine learning techniques that harness signals from movement or speech data as objective indicators of cognitive and neuropsychological functioning with the aim to shorten the long diagnostic process. Specifically, we investigate the potential of Path Signatures as a mathematical tool to quantify and extract meaningful features from video and speech data. Path Signatures provide a robust mathematical framework capable of capturing intricate temporal dependencies and patterns in sequential data. An extensive review in \cite{lyons2022signature} outlines their application in estimating gesture correctness, predicting bipolar disorder, Alzheimer's disease, or early sepsis, as well as in decision-making, speech emotion recognition, and analysis of medical prescriptions.  

The ultimate goal is to streamline the diagnostic process by reducing the quantity and duration of traditional assessments, leading to more efficient evaluations and faster access to appropriate interventions. In doing so, we can promote equity and improve access by democratizing families' ability to access and engage in timely and efficient clinical diagnosis\cite{buch2018artificial}. 

In the subsequent sections of this research report, we present our dataset, pre-processing, the application of ML tools to quantify movement and speech from which we then compute Path Signatures. We explore correlations between these features and target cognitive assessment scales and evaluate Support Vector Machine (SVM) models that attempt to classify children's cognitive assessment outcomes based on the extracted features.  Finally, we discuss the potential of Path Signatures reflective of patient-child interaction dynamics for the shortening of cognitive and neuropsychological assessments.

\section{Experimental Setting and Dataset}

\subsection{Dataset}

This research report presents a post-hoc analysis of recordings of routine clinical assessments where the patients and their legal guardians provided written, informed consent/assent to their being recorded and the data being used for clinical analysis and further research. The dataset is composed of 39 ADHD child-clinician interactions during the administration of neuropsychological tests. In terms of demographics, participants are all US-English speakers with an average age of 10 years, and a standard deviation of 3 years, indicating a relatively homogeneous cohort. The gender distribution of the group is slightly skewed towards males, who make up about $63\%$ of the participants. In terms of the recording duration, the total length of the clinical interviews per participant ranges between 5 to 6 hours. 

\subsection{Clinical Sessions Outline}

All participants underwent a routine, clinical diagnostic assessment with one of two clinicians. These sessions involved a clinical interview of the patient (and in some cases informant interview - e.g. parent interview), as well as the standardized administration of standardised neuropsychological tests which included:  
\begin{itemize}

\item Wechsler Intelligence Scale for Children, Fifth Edition (WISC-V): A cognitive assessment tool for children aged 6 to 16. It measures intelligence using 10 subtests and calculates composite scores, among which we selected the Full Scale IQ as a measure of general intelligence. 

\item Test of Everyday Attention for Children (TEA-Ch): Measures attention abilities in daily life situations. It consists of eight subtests assessing selective, sustained, and divided attention. Access to the score ranges, narrowed our focus on the Sustained Attention, which assesses the ability to hold attention on a "boring" yet important task.

\item Clinical Evaluation of Language Fundamentals, Fifth Edition (CELF-5): Evaluates language skills, including receptive and expressive abilities. It consists of subtests assessing semantics, syntax, and pragmatics. Similarly to above, we used the composite scores for Formulated Sentences going forward, as we had access to score ranges for this scale. It assesses the ability to formulate complete, semantically and grammatically correct sentences, reflecting the capacity to integrate semantic,
syntactic, and pragmatic rules and constraints while using working memory.

\item A Developmental NEuroPSYchological Assessment, Second Edition (NEPSY-II): Measures neuropsychological abilities in children and adolescents aged 3 to 16. Due to having access to the score ranges, we used the subtest of social cognition: Affect Recognition, assessing the ability to decode emotion expressions.

\end{itemize}

All neuropsychological measures were scored using standardized normative samples as provided by each test publisher. Specifically, the Wechsler Intelligence Scale for Children (WISC) results span from a minimum of 40 to a maximum of 160, with an average score of 100 and a standard deviation of 15. The other three scales in use exhibit a range from 1 to 19, with a mean score of 10 and a standard deviation of 3. 

Raw scores were converted to demographically-adjusted standard scores and percentiles. The compilation of these standardized scores comprised each individual's neuropsychological profile, with objective data points mapping individual cognitive domains (e.g. attention, memory).

 Table \ref{table:scores} presents a summary of the distribution of pertinent scores. In this table, we categorize scores as 'Low', 'Medium', or 'High'. 'Low' represents those scores that are one standard deviation below the benchmark population, 'High' designates scores that exceed one standard deviation above the benchmark, and 'Medium' refers to all other scores.

\begin{table}[h!]
\caption{Distribution of test results}
\centering
\begin{tabular}{|c|c|c|c|}
\hline
\textbf{Test} & \textbf{Low} & \textbf{Medium} & \textbf{High} \\
\hline
\hline
 $WISC$ & $6\%$ & $58\%$ & $36\%$  \\
\hline
 $TEA$ & $60\%$ & $20\%$ & $20\%$  \\
\hline
 $NEPSY$ & $3\%$ & $58\%$ & $39\%$  \\
\hline
 $CELF$ & $3\%$ & $44\%$ & $53\%$  \\
\hline
\end{tabular}
\label{table:scores}
\end{table}
\section{Methods}
\subsection{Pre-processing}

To investigate patient-clinician interactions and their potential implications for diagnoses like ADHD and ASD, we implemented a pre-processing pipeline that leverages various ML techniques and libraries on our dataset of video recordings.

First, the recorded videos had various technical specifications, including .mov and .mp4 formats and ranging from audio sampling rates of 48 KHz to 8 KHz. For standardisation purposes, we extracted the audio from each recording and stored it in a separate wav file at 16khz (downsampling or upsampling as necessary).
The next step was to segment the videos according to the particular tests being conducted. This was accomplished using a speech-to-text model, specifically the Whisper model \cite{radford2023robust}, to transcribe clinician instructions into text. The text was then matched with known test instructions using text similarity scores, effectively labeling segments of the video according to the tests performed.
In the process of segmenting the recordings, it was found that the WISC test consistently spanned the initial 40 minutes of dialogue between child and clinician. As it represents a standardised test with predictable structure, we decided to select this window of the first 40 minutes of each recording, in order to ensure equivalent video and speech information across the different child-clinician sessions.
In the next phase, we examined the dynamics of conversation through the identification of \textit{speech turns}. To distinguish between the voices of the patient and the clinician we used Hugging Face's Pyannote, a state-of-the-art pre-trained speaker diarization model \cite{Bredin2020, Bredin2021}. This process allowed us to attribute each turn in the conversation to either the patient or clinician, thus offering insights into their dialogue patterns.
The final phase of our pre-processing focused on movement tracking within the video segments. The sessions were captured in a range of clinical settings, hence, the lighting and seating conditions were quite varied and occasionally involved the presence of more than two individuals (e.g. parent or additional clinic personnel). Recognizing the complexity of this task, we focused on head movements. To this end, we made use of OpenVINO's pre-trained model, specifically human-pose-estimation-0001 from the OpenVINO Model Zoo by Intel. 

\subsection{Features Extraction}
After pre-processing, our dataset is constituted by three data subsets for each recording: 
\begin{itemize}
    \item  head positions of the patient and clinician as a time series of 2D coordinates; 
    \item timestamps denoting speech turns of the clinician and patient;
    \item the numerical scores from all performed tests.
\end{itemize}

The high-dimensionality of the data presents a challenge due to the relatively limited number of subjects in the dataset. Thus, the task is to find a compact but expressive set of features that can effectively reduce the problem's dimensionality while maintaining data richness.

In addition to computing standard summary statistics, we identified Path Signatures as a promising tool to succinctly represent the data. Given the small size of our dataset, Path Signatures \cite{chevyrev2016primer}, with their demonstrated competitiveness with state-of-the-art deep learning approaches in similar scenarios \cite{bonnier2019deep}, make a compelling choice for feature extraction.

\subsubsection{Path Signatures}

The central concepts and definitions of path signatures are as follows:

Given d-dimensional data streams, $x(t)$, represented as discrete time series of N elements:

\begin{equation}
    x(t) =\begin{pmatrix} x^{(1)}(t) \\ . \\ . \\ . \\ x^{(d)}(t) \end{pmatrix}, \;\;\; x_i =\begin{pmatrix} x^{(1)}(t_i) \\ . \\ . \\ . \\ x^{(d)}(t_i) \end{pmatrix}
\end{equation}
for $i = 1...N$.

Each stream is perceived as a path in d dimensions. Path signatures are constructed as iterated integrals of the path, capturing non-linear features of the data. The first-order signature $S^{(i)}(x)_{ab} = \int_{a}^{b} dt \dot{x}^{(i)}(t) = x^{(i)}(b) - x^{(i)}(a)$ is merely the difference between the final and initial values of the series. Higher-order signatures are defined iteratively and are labelled by the sequence of coordinates along which the integrals are computed. The all-order signature is expanded as a formal power series over the basis of monomials $\mathbf{e_{i_k}}$, with $k=1\dots d$, which identify the $d$-dimensions in which the path is embedded. The indexing set of the terms of the signature of a path is the collection of all multi-indexes ${(i_1\dots i_k)}$. The symbol $\otimes$ is the tensor product between monomials, see also \cite{chevyrev2016primer}.

\begin{equation}
\label{eq:sig_pow_ser}
    S(x)_{ab} = \sum_k\,\sum_{i_1\dots i_k} S^{(i_1,\dots, i_k)}(x)_{ab}\,\mathbf{e_{i_1}}\otimes\dots\otimes\mathbf{e_{i_k}}.
\end{equation}

Signatures possess several mathematical features that render them ideal for path description:
\begin{itemize}
    \item Signatures uniquely identify the path.
\item They record the sequence of events regardless of how time is parameterized. For any reparameterisation $t \rightarrow \psi(t)$, with $\psi(t)$ never decreasing with $t$ and maintaining $\psi(a) = a$ and $\psi(b) = b$, it always holds that $S_{a,b}(x(t)) = S_{a,b}(x(\psi(t))$.
\item High-order signatures decay rapidly, implying in theory that it's unnecessary to compute very high order signatures. However, in practice, this may be true only for extremely high orders.
\item Universal non-linearity allows learning any function (even non-linear) on the path as a combination of signatures up to a fixed order. Although this is insightful, it does not provide a practical limit to the required order.
\item The log-signature contains all the essential information in the minimal number of components by considering all the algebraic relations between signature components. The log-signature is written to all orders as a power series over the Lie brackets $\left[\mathbf{e_{i_1}},\mathbf{e_{i_2}}\right] = \mathbf{e_{i_1}}\otimes\mathbf{e_{i_2}} - \mathbf{e_{i_2}}\otimes\mathbf{e_{i_1}}$
\begin{equation}
    \log S(x)_{ab} = \sum_k\!\sum_{i_1\dots i_k}S^{[i_1,[i_2,[\dots, i_k]]]}[\mathbf{e_{i_1}},[\mathbf{e_{i_2}},[\dots,\mathbf{e_{i_k}}]]],
    \label{log-sig}
\end{equation}
where the components are determined in terms of those in eq.~(\ref{eq:sig_pow_ser}), for instance $S^{[1,2]}(x)_{ab} = 1/2(S^{(1,2)}(x)_{ab}-S^{(2,1)}(x)_{ab})$. In the rest of the research report we omit the dependence on the path and on the endpoints points $a$ and $b$ to simplify notations.
\end{itemize}

The way we embed time series data into a path is of great importance in computing path signatures. For single-dimensional data, it's critical to represent it in a higher-dimensional path to avoid limiting the signature to merely reflecting the total variation in the series. For multi-dimensional data, various embeddings can capture different types of information, with transformations such as Time Augmentation and Lead-Lag transform being the most notable ones.

The transformation applied to the data before computing the signature dramatically influences the output. As we always truncate the signature at a certain order, some path information inevitably gets omitted. If the information we seek relies on these higher-order signatures, the truncation might hinder our analysis. However, each order in the signature encodes different data, depending on the transformation applied to the stream. For example, using Time Augmentation necessitates a second-order signature to get the mean of the time series. Conversely, combining the Cumulative Sum and Lead-Lag transform allows us to extract the mean and standard deviation from the first and second order in the signature, respectively. For more information on this we refer to \cite{lyons2022signature}.

In the subsequent sections, we will detail the methodology used to transform the available data into paths that can be processed using path signatures.

\subsubsection{Speech Turns Paths}
\label{speech-feat-section}
\begin{figure}[h]
\centering
\includegraphics[scale=0.5]{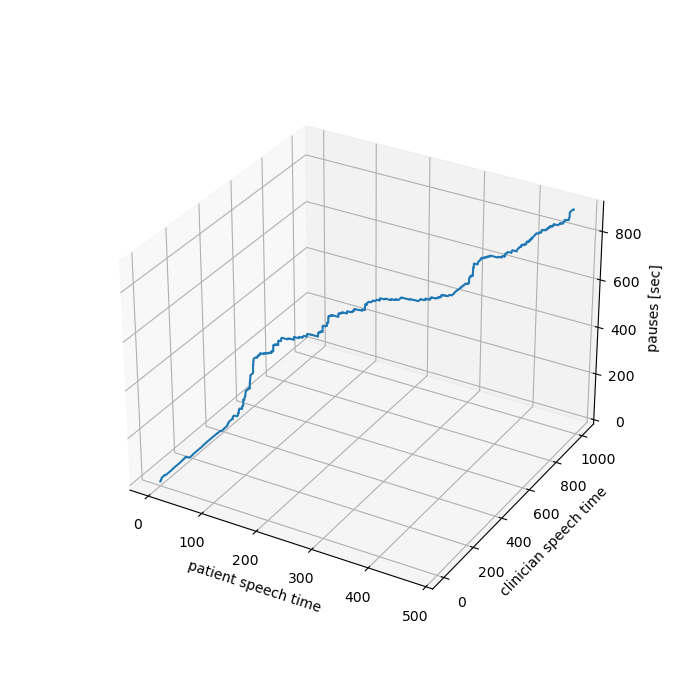}
\caption{Example of a speech-turns path representing pauses, clinician speech, and patient speech.}
\label{fig-turns}
\end{figure}

The dynamic interactions within the interviews are represented as a sequence of speech turns, alternating between the patient, clinician, and periods of silence. These elements can be regarded as distinct dimensions, enabling us to depict the unfolding conversation within a three-dimensional space, as illustrated in Figure \ref{fig-turns}.

From each path, we extract fundamental features such as:

\begin{itemize}
    \item The absolute and relative ($\%$) count of turns taken by the patient ($p_{\text{cnt}}$ and $p_{\text{crel}}$, respectively), clinician ($c_{\text{crel}}$ and $c_{\text{crel}}$, respectively), and pauses ($s_{\text{cnt}}$ and $s_{\text{crel}}$, respectively).
    \item The cumulative speech duration of the patient and clinician, as well as total silence, both quantified in seconds ($p_t$, $c_t$ and $s_t$, respectively) and as a percentage of total time ($p_r$, $c_r$ and $s_r$, respectively).
    \item The mean and variance of the speech turns for both the patient and clinician ($p_{\text{mean}}$, $p_{\text{std}}$ and $c_{\text{mean}}$, $c_{\text{std}}$, respectively).
    \item Additionally, we compute the log-signatures up to the 4th order as defined in Eq. \ref{log-sig} using the software \texttt{esig} \cite{esig}. To simplify notations log-signatures are labelled only with their component monomials, i.e. $p_t \equiv S^{(p_t)}$ is the first-order signature in the component $p_t$ and $[p_t,c_t] \equiv \frac{1}{2}(S^{(p_t,c_t)}-S^{(c_t,p_t)})$.
    These metrics uniquely represent the path and encode the sequence of events.  
\end{itemize}

In total, we utilize 46 unique features for our analysis.
This approach provides key insights into conversational dynamics, such as patient responses following clinician remarks: Are their responses immediate or delayed? Are there pauses? Are the responses lengthy?
\subsubsection{Patient-Clinician Movements Paths}
\label{video-feat-section}
We track the head movements of the patient and clinician throughout 10,000 frames of video (note: all videos have a frame rate of 15 fps). Similar to the speech features, we generate 3-dimensional paths. 

\begin{figure}[h]
\centering
\includegraphics[scale=0.5]{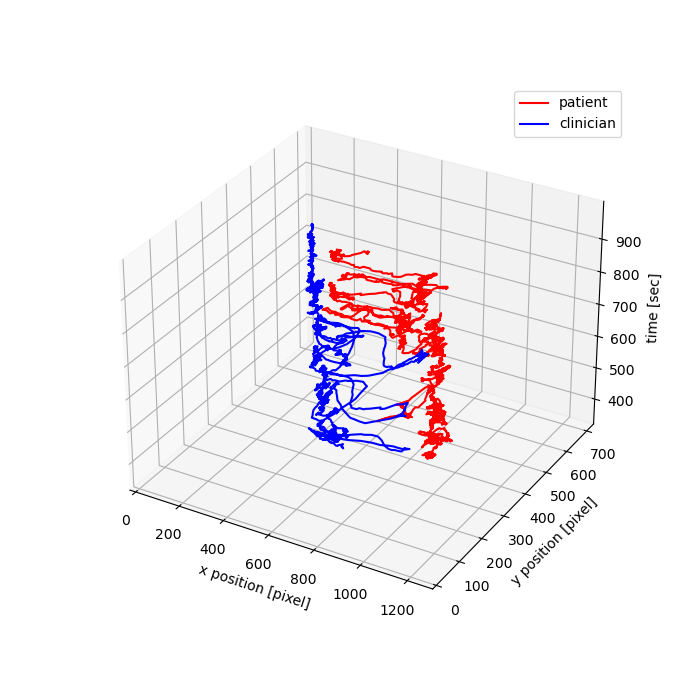}
\caption{Example of clinician and patient head movements paths.}
\label{fig-movement}
\end{figure}

We compute a total of 41 video features, categorized as follows:

\begin{itemize}
    \item The standard deviation of the patient's position in pixels, $x_{p,\text{std}}$ and $y_{p,\text{std}}$(note that the mean position isn't as relevant, as it mostly indicates the placement of the chair).
    \item Level 3 path log-signatures extracted from 3-dimensional paths corresponding to the (x, y, time) coordinates of the patient, $(x_p(t), y_p(t), t)$.
    \item Level 3 path log-signatures extracted from the 4-dimensional path that includes (x-patient, y-patient, x-clinician, y-clinician), i.e. $(x_p,y_p,x_c,y_c)$, allowing us to capture correlations between the patient's and clinician's actions.
\end{itemize}

The speed of head movements, a key aspect mentioned in \cite{7961818}, hasn't been tracked in our current model. Although speed can be derived from the existing paths by computing the difference in head positions between frames, this approach fails to account for varying camera-to-patient distances across videos and would require further pre-processing work. 

The constructed path signature features are inherently robust to these issues. They remain constant when a constant shift is added to the path or when time parameters are modified, except when time itself is a component of the path. Importantly, path signatures retain the information on the order of events, making them effective in encoding patient-clinician synchronicity, a key element highlighted in \cite{celiktutan2022computational} when studying autistic behaviors.

\section{Experiments and Results}

We are investigating the hypothesis that the nature of patient-clinician interactions might hold predictive power for the results of the clinical assessments. In the upcoming section, we first delve into a focused correlation analysis between the extracted features that encapsulate patient-clinician interactions and the outcomes of four cognitive assessment tests conducted during the diagnostic sessions. We consider features from both audio and video data, seeking potential patterns that could correspond with the clinical test results. While each feature's correlation will be examined individually, we also explore combinations of features for potentially stronger correlations. The end goal is to uncover any latent predictive capabilities of these interaction dynamics for clinical outcomes.

Given our dataset's limited size, we utilize bootstrapping to ascertain the significance of these correlations.We sampled our dataset with replacement and recalculated the correlations, thereby generating a distribution for each correlation. This process enabled us to estimate the variability of our correlation coefficients and help discern which correlations were significant, thus providing a more robust understanding of the relationships between our features and the target scales.

In the following section, we use SVM models to explore whether it is possible to classify the outcome of each of the cognitive assessment tests using those features. We conduct a four-fold cross-validation for validation and explore the confounding effect of demographics variables.   

\subsection{Correlation Analysis}

The exploratory analysis of the extracted features and the target scale commenced with a correlation analysis. This preliminary analysis provided us with insights into the relationships between the features and the target scale and within the feature set itself. Through this correlation analysis, we were able to discern patterns of co-variation, potential redundancies, and dependencies among the features.

We labeled the features in $4$ groups for readability and to simplify the discussion.
The first group, labeled as $speech\_stats\_0, ..., speech\_stats\_16$, represents simple statistical features derived from speech, which include total pause duration, mean time of speech for the patient, and other related metrics as described in Section \ref{speech-feat-section}. 

The second group, $speech\_path\_0, ..., speech\_path\_28$, are the log-signatures on speech paths as described in \ref{speech-feat-section}. 

The third group, $video\_stats\_0, ...., video\_stats\_8$, refers to simple statistical features extracted from the movement paths as described in Section \ref{video-feat-section}.

Lastly, the group $video\_path\_0, ..., video\_path\_36$ encompasses the log-signature features derived from the movement paths.

\begin{figure}[h]
\centering
\includegraphics[scale=0.4]{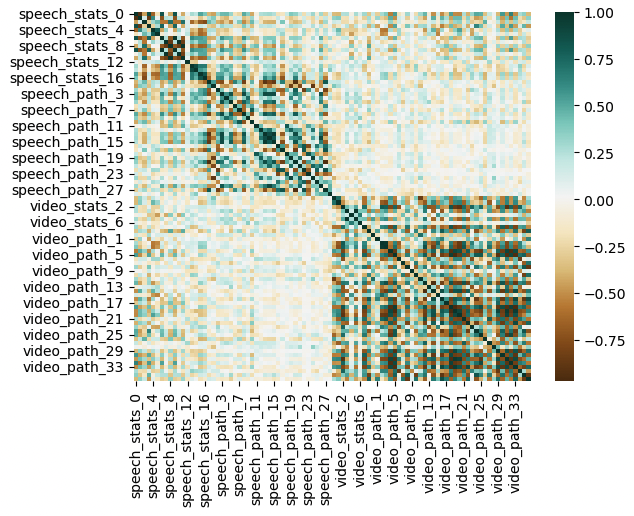}
\caption{Correlation matrix of the speech and video features. Brown to green represent negative to positive Spearman correlation coefficient values.}
\label{fig:features:correlation}
\end{figure}

Looking at Figure \ref{fig:features:correlation} we can see that:

\begin{itemize}
    \item There are three groups of features that appear to be orthogonal: simple statistical features on speech, path signatures on speech and video features.
    \item The path signatures on speech feature group shows some redundancy, but also contains several independent features, potentially contributing unique insights into speech patterns.
    \item  The video features group exhibits a high degree of intra-correlation, suggesting redundancy but also hinting at consistent patterns. Future analysis may aim to distill the most informative, non-redundant features within this group.
\end{itemize}
 
\begin{figure}[h]
\centering
\includegraphics[scale=0.25]{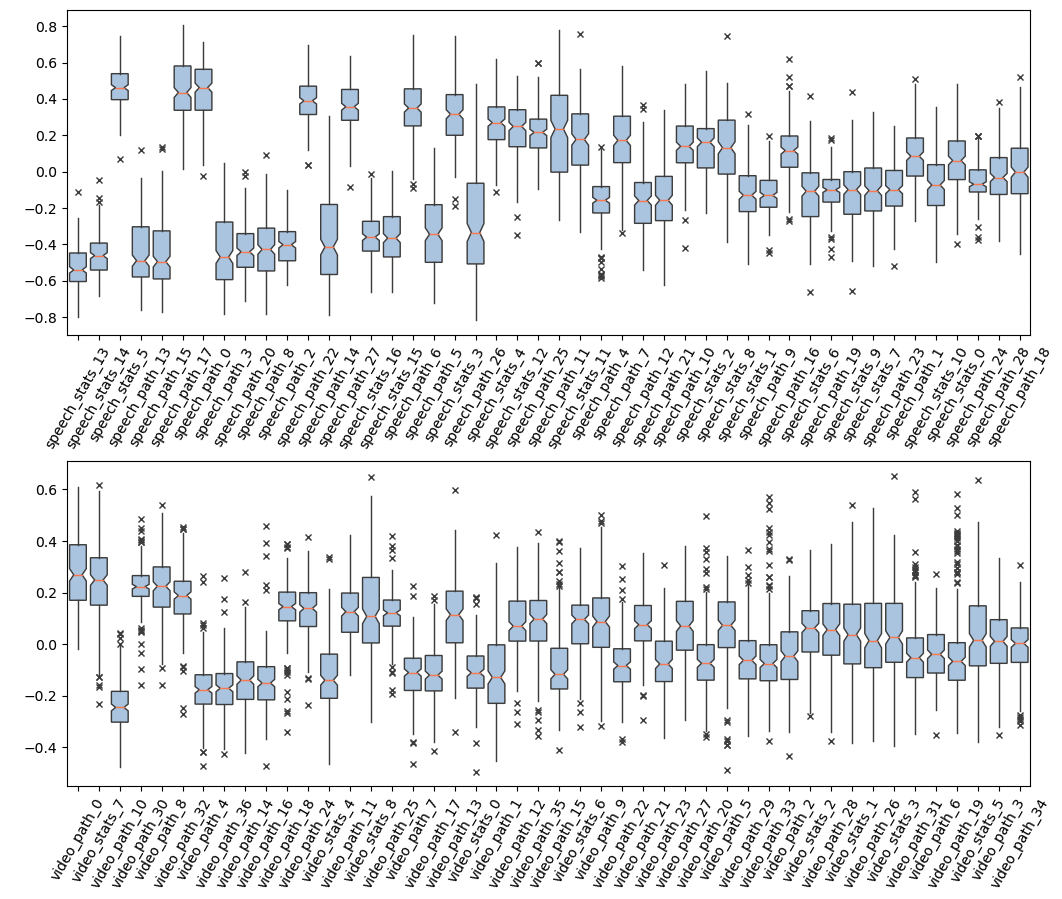}
\caption{Bootstrapped correlation between features and the CELF-5 Formulated Sentences scale (selected for illustration purposes). Y-axis represents bootstrapped Spearman correlation coefficients. In the top panel we are showing all the speech related features, while in the bottom panel we display the video features.}
\label{fig:celf:correlation}
\end{figure}

Progressing in our analysis, we then examined the correlation between the extracted features and the target scales. We present the full list of correlations between the CELF-5 Formulated Sentences scale and the extracted features in Figure \ref{fig:celf:correlation}. The selection of the CELF-5 Formulated Sentences as a scale to display our results was random - it is used for illustration purposes (data on the other scales are in table form).  To facilitate readability, the features have been ordered by their absolute mean correlation value. Upon examination, it is evident that a substantial number of features demonstrate significant correlation with the CELF-5 Formulated Sentences scale. Remarkably, a majority of these significant correlations are associated with the path signature features, several of which are of high order. This provides preliminary support for the potential utility of these high-order path signature features in cognitive assessment models.

As the figure shows, bootstrap correlation analyses for the other scales yield similar results, with average correlations reaching as high as 0.6. This consistent pattern across different scales strengthens the validity of our findings and underlines the potential of our features to predict various cognitive assessment outcomes.
 
\subsection{Predictive Models Results}
In the next step, we frame the predictive task as a binary classification problem across four cognitive measures, namely, WISC-v Full-Scale IQ, NEPSY-II Affect Recognition, CELF-5 Formulated Sentences, and TEA-Ch Sustained Attention. Given the distribution of data across these measures (see Table \ref{table:scores}), in order to achieve a sensible binary split, we categorize cases labeled with 'Low' and 'Medium' scores as 'True' for WISC, NEPSY-II, and CELF-5. Conversely, for TEA-Ch, we only consider the 'Low' labeled cases as 'True'. Recognizing the orthogonality between speech and movement features as seen in Figure \ref{fig:features:correlation}, we integrate all of them as a feature set into our predictive model, leveraging the potential for complementarity and nuanced interpretation.

In terms of machine learning methodology, we utilize an SVM model with a linear kernel. This choice is motivated by the model's robustness and its capacity to handle high-dimensional data \cite{wu2008kernel}, as it's the case with our feature set combining speech and movement features. We employ an L2 regularization scheme in the SVM model to prevent overfitting and to ensure the generalizability of our results \cite{chang2011libsvm}, \cite{scikit-learn}.

To evaluate the performance of our model and ensure its robustness, we conduct a four-fold cross-validation. This strategy ensures that our findings are not due to a specific random split of the data but are indeed inherent to the data and the model's ability to generalize. Performance is reported in terms of the Area Under the Receiver Operating Characteristic Curve (AUC), providing an effective summary of the classifier performance across all possible classification thresholds. 

\begin{figure}[h]
\centering
\includegraphics[scale=0.45]{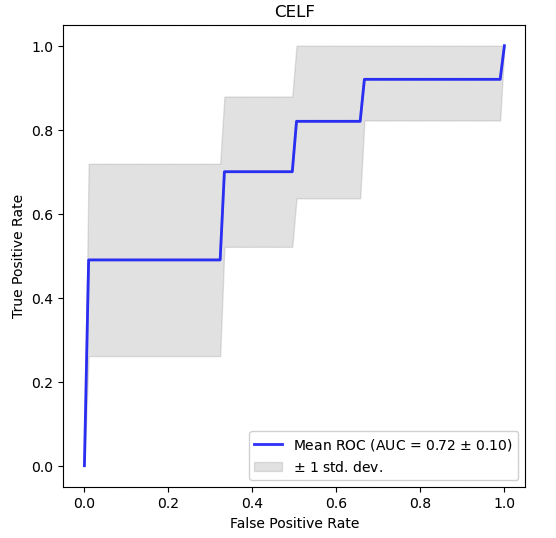}
\caption{Cross-validation ROC plot of Support Vector Machine classifier for the CELF-5 (Formulates Sentences subscale) using speech turns and movement features.}
\label{fig-classification-model}
\end{figure}

Figure \ref{fig-classification-model} displays the ROC curve for one target scale, visually depicting our model's performance. Conversely, Table \ref{table:AUC:model} enumerates the mean and standard deviation of the AUC scores computed on the test folds for all target measures, giving an overall view of both the performance and the variability of the model across different data subsets.

\begin{table}[h!]
\caption{Cross-validated AUC mean and standard deviation for the binary classification models}
\centering
\begin{tabular}{|c|c|c|}
\hline
\textbf{Scale} & \textbf{AUC mean} & \textbf{AUC Std}  \\
\hline
 $WISC$ & 0.71 & 0.18 \\
\hline
$CELF$ & 0.72 & 0.10 \\
\hline
$TEA$ & 0.87 & 0.14  \\
\hline
$NEPSY$ & 0.77 & 0.09  \\
\hline
\end{tabular}
\label{table:AUC:model}
\end{table}

In an effort to understand whether demographic factors serve as confounding variables or skew the results, we can compare the performance supplementary SVM models utilizing age and gender as input features with our original models. The results are presented in Table \ref{table:AUC:demographics}. 

\begin{table}[h!]
\caption{Cross-validated AUC mean and standard deviation for the binary classification task using demographics data as features}
\centering
\begin{tabular}{|c|c|c|}
\hline
\textbf{Scale} & \textbf{AUC mean} & \textbf{AUC Std}  \\
\hline
 $WISC$ & 0.70 & 0.13 \\
\hline
$CELF$ & 0.41 & 0.14 \\
\hline
$TEA$ & 0.41 & 0.26  \\
\hline
$NEPSY$ & 0.61 & 0.26  \\
\hline
\end{tabular}
\label{table:AUC:demographics}
\end{table}

The classification models constructed using speech turn and movement features consistently demonstrate AUC values significantly above chance level in cross-validation, underscoring the predictive efficacy of these features. On the other hand, when we conduct similar exercises utilizing demographic data solely as input, we observe a discernible predictive ability above chance level only for the WISC-V test, which confirms clinical observations and could potentially be interpreted as an expression of the detrimental effects of delayed diagnosis on cognitive. The AUC scores for all other scales approximate 0.5, suggesting a performance consistent with chance, suggesting that demographics variables aren't important factors for predicting the target classes. 

These findings offer preliminary evidence that the features that characterize clinician-patient interactions, from speech turns and movements in the initial 40 minutes of recordings isn't limited to the WISC-V tests from the recording of which we derived the features, but may hold predictive value for subsequent test outcomes, suggesting the potential for such models to expedite the clinical session and predict later scores—potentially saving several hours.

\section{Conclusions and Discussion}
In this study we used machine learning tools to extract speech and movement features from recordings of cognitive assessments of children with ADHD. We used the path signature method to quantify novel speech and movement metrics specifically describing the interactive dynamic between clinician and patient, as well as to  demonstrate the feasibility and utility of these metrics to both correlate with and predict neuropsychological performance. Specifically, we found that speech turn patterns between patient and clinician were shown to significantly correlate with objective cognitive outcome metrics as assessed via standardized clinical neuropsychological measures commonly used in clinical settings (e.g. WISC-5).  Similarly, the relationship between the patient and the clinician’s movement was a strong predictor of neuropsychological performance.  Importantly, not only did these metrics correlate with neurocognitive outcomes, but the speech and movement metrics from the first 40 minutes of audio-video data significantly predicted the neurocognitive outcome data from the remaining cognitive tests and several hours of neuropsychological evaluation sessions. While preliminary, this could offer a significant advancement over previous methods and expands the possibilities of using such metrics for predicting both clinical and real world neurocognitive outcomes using much less data and much less time than a full neuropsychological evaluation requires. By predicting neuropsychological test performance across an entire battery of tests, and therefore across all neurocognitive domains (e.g. attention, processing speed, learning, memory), we demonstrate the robust utility of these speech and motion metrics to predict a patient’s full neurocognitive profile. If used clinically, this could provide rich clinical data in a much shorter and less costly time than is required for a full neuropsychological evaluation. This could also serve as a screening tool for determining which patients may or may not need to undergo a full neuropsychological evaluation, thereby reducing the burden on patients, schools and the overall healthcare system.
This helps to democratize access and reduces the negative impact of bottlenecks on patients, allowing them access to evaluations and care sooner than is currently possible and reducing the deleterious impact of their condition on educational, occupational, and social functioning as well as overall wellbeing and quality of life. Further, neuropsychological evaluations are highly impacted by the limitations of the language the tests are administered in and the cultural context of the test items \cite{fernandez2018bias}. This reduces the access to these tests further and limits the utility with patients whose native language is different from that of the test. By generating objective metrics that are agnostic to language and cultural context from speech and movement signals, this tool further expands the range of people with access to neurocognitive evaluations and subsequent care.  
Limitations of this study include a small sample size which was addressed statistically through bootstrapping and will be further addressed in future work by both the amassing of a larger clinical data set as well as the expansion of speech and motion metrics to include more fine grained analyses such as speech prosody, facial microexpressions, and the concordance, coherence, and/or lack thereof between speech, facial and body movement metrics. Additionally, while it is meaningful to consider interpersonal dynamics in the diagnostic process \cite{georgescu2019machine},  \cite{koehler2022brief}, \cite{celiktutan2022computational} future work may seek to develop patient-only or intrapersonal dynamics (e.g. coordination between speech and movement within an individual) in order to remove the requirement for clinicians to be included in the dataset. This would expand the range of use cases. To this end, while variability of physical location was not a significant metric in this analysis, future analyses could attempt to improve the utility of this and other such metrics through normalization procedures. Taken together, the inclusion of a larger sample size, more fine-grained metrics and additional analytic tools will allow for more successful amelioration of the effect of outliers on our model and results in the future. Finally, future models should be expanded to include non-English, non-American culture patients in order to enhance the model to ensure it is agnostic to language and culture. This will further democratize access to neuropsychological evaluations.
Taken together, this study demonstrates the initial feasibility and clinical utility of artificial intelligence-based metrics of speech and motion for predicting neurocognitive performance in patients undergoing neuropsychological evaluations. By further refining this procedure, such a tool could be used in the future to democratize access to neuropsychological evaluation, reduce burden on school and healthcare systems, and improve patient care and outcomes.

\section*{Acknowledgment}

We would like to thank Dr. Nicholas Cummins for his helpful feedback with an earlier version of this manuscript.

\bibliographystyle{IEEEtran}
\bibliography{bibliography}

\end{document}